\documentclass[review]{elsarticle}
\usepackage{amsmath}
\usepackage{lineno,hyperref}
\usepackage{algorithm}
\usepackage{algpseudocode}
\usepackage{xcolor}
\usepackage{pxfonts}
\usepackage{comment}
\usepackage{graphicx} 
\usepackage{booktabs} 
\usepackage{caption}
\usepackage{subcaption}
\usepackage[capitalise]{cleveref}

\modulolinenumbers[5]

\begin{document}

\begin{frontmatter}

\title{Longest Common Subsequence: Tabular vs. Closed-Form Equation Computation of Subsequence Probability}

\author[mymainaddress]{Alireza Abdi}
\ead{alirezaabdi@iasbs.ac.ir}

\author[mymainaddress]{Mohsen Hooshmand\corref{mycorrespondingauthor}}
\cortext[mycorrespondingauthor]{Corresponding author}
\ead{mohsen.hooshmand@iasbs.ac.ir}

\address[mymainaddress]{Department of Computer Science and Information Technology,\\ Institute for Advanced Studies in Basic Sciences (IASBS), Zanjan, Iran}

\begin{abstract}
The Longest Common Subsequence Problem (LCS) deals with finding the longest subsequence among a given set of strings. The LCS problem is an NP-hard problem which makes it a target for lots of effort to find a better solution with heuristics methods. The baseline for most famous heuristics functions is a tabular random, probabilistic approach. This approach approximates the length of the LCS in each iteration. The combination of beam search and tabular probabilistic-based heuristics has led to a large number of proposals and achievements in algorithms for solving the LCS problem. In this work, we introduce a closed-form equation of the probabilistic table calculation for the first time. Moreover, we present other corresponding forms of the closed-form equation and prove all of them. The closed-form equation opens new ways for analysis and further approximations. Using the theorems and beam search, we propose an analytic method for estimating the length of the LCS of the remaining subsequence. Furthermore, we present another heuristic function based on the Coefficient of Variation. The results show that our proposed methods outperform the state-of-the-art methods on the LCS problem.

\end{abstract}

\begin{keyword}
\texttt{Longest Common Subsequence Problem, Beam Search, Beta Distribution}. 

\end{keyword}

\end{frontmatter}

\section{Introduction}
The Longest Common Subsequence Problem (LCS) is a classic problem in the theoretical domain of computer science, such as algorithm design and complexity~\cite{Wagner74, jiang95}. However, it is not limited to a theoretical problem. It plays an important role in a wide range of fields, e.g., sequence alignment and pattern discovery in the area of Bioinformatics~\cite{shikder19}, graph learning~\cite{huang21}, trajectories matching and similarity~\cite{dong18, Ding19}. The LCS is about having several strings over a finite alphabet. For example, DNA sequencing has four bases (four characters) of A, G, C, and T \cite{smith81}. The goal of the LCS problem is to find the longest possible subsequence among a given set of strings. The ``Subsequence'' term in the LCS means the importance of order. Having two strings in the given set is the simplest case of the LCS problem. We call it the ``2-LCS'' problem. Solving the 2-LCS is optimally and practically possible~\cite{gusfield97}. There are plenty of methods to solve the 2-LCS problem, like integer programming~\cite{singireddy03}, dynamic programming~\cite{sankoff72, smith81,hirschberg77, apostolico92,masek80}, and branch and bound~\cite{Easton08}. The time complexity of dynamic programming is $O(l^{N}).$ Where $N$ is the number of strings and $l$ is the length of the longest string in the given set~\cite{smith81}. However, the plot will be different when the number of strings increases to more than two strings. In the same way, we call solving the LCS of the set of strings with cardinally greater than two as ``M-LCS''. In the rest of this paper, we use LCS instead of M-LCS. Maier showed that the LCS, in general, is NP-hard \cite{maier78}. Thus, finding the LCS problem's optimal solution is generally impossible. Subsequently, many efforts have been made to find the near-optimal solution of the LCS problem with approximation~\cite{bonizzoni01approximation,chin94approximation,jiang95approximation} and heuristic~\cite{blum09,blum10, mousavi12improved,djukanovic19} methods. Blum et al. \cite{blum09} in their seminal work, proposed the beam search framework to find the near-optimal solution to the LCS problem heuristically. Later, Mousavi and Tabataba used a randomized tabular probabilistic heuristic, as well as the power heuristic on beam search \cite{mousavi12improved, tabataba12hyper}. Blum, in collaboration with other researchers, has continued proposing new heuristics and approximations to improve the quality of solutions of the LCS problem~\cite{djukanovic19}. From that onwards, the probabilistic tabular computation has become the basis of beam search framework~\cite{djukanovic19}. It is worth mentioning that other search strategies were employed to solve the LCS problem. $A^{*}$ is one of them. In combination with $A^{*}$ and column search, Djukanovic et al.~\cite{djukanovic20Astar} proposed a new anytime $A^{*}$ algorithm that could reach the highest quality of solutions. As Djukanovic et al. mentioned in their work~\cite{djukanovic20Astar}, the success of their method is due to the use of the $BS$-$Ex$ heuristic function, which was introduced in~\cite{djukanovic19}. However, the weight of memory and time is heavier than the mentioned heuristic in the success of their algorithm. This algorithm belongs to the set of ``long-run'' algorithms, which their solution depends on the available memory and time interval allowed for running. Our proposed methods in this paper belong to the ``short-run'' algorithms, which rely on the quality of the heuristic rather than available resources.

In this work, as the main contribution, we present several closed-form equations for the probabilistic tabular computation for the first time and prove them. As another representation of the closed form equation, we show that it is nothing but a weighted sum of $Beta$ distribution functions. In each step of the algorithms on the LCS, some scores are computed, and those candidates with the best values are chosen for expansion and selection. These scores depend on some parameters, which we define in the following sections. Before that, these parameters were computed with no theoretical or practical background. As another contribution of this work, we convert the calculation of such parameters from blind estimation into an analytical form. Last but not least, we introduce two methods by utilizing the power of the proposed theorems and the behavior of strings. The proposed methods outperform the state-of-the-art methods.

In section~\ref{req}, we define the requirements of our work. In section~\ref{rewo}, we describe the related work. In section \ref{equat} we introduce theorems and corollaries in the equational format of probability computation for the first time.
Section \ref{method} devotes to the analysis of a new way of estimating the best remaining length of strings. Moreover, we propose another heuristic derived from the Coefficient of Variation.
Experimental results are reported and discussed in section \ref{results}. Finally, section \ref{conc} concludes the paper. Also, we provide proofs of proposed theorems as well as some extra results in Appendices which come at the end.  

\section{System model}\label{req}
In this section, we go through the basic concepts and definitions necessary for the proper formulation of the problem. First, we explore notations and mathematical requirements. Then, we briefly define the beam search and show its use in the LCS problem.  
\subsection{Notations and preliminaries}
Let $\Sigma = \{ \sigma_{1}, \sigma_{2}, \cdot \cdot \cdot, \sigma_m\}$ be the alphabet set of input strings and the number of alphabet is $m$, $|\Sigma|=m$. The set $S = \{s_{1}, s_{2}, \cdots, s_{N} \}$ is those strings which we aim to find the longest common subsequence among them. The symbol $N$ is number of strings, or equivalently $|S|=N$.  In the constructive solving of the LCS problem, in each step, a character from $\Sigma$ is chosen as the part of the LCS, called $\nu$. Then, the character $\nu$ and all prior characters of $\nu$ are removed from each string $s_i, i\in\{1,\cdots,N\}$. The remaining, or the postfix after $\nu$, is shown by $r_{i}^{\nu}$. Putting postfixes together in a set, $R^{\nu}$ is the set of all $r_{i}^{\nu}$-s, $i\in\{1,\cdots,N\}$. 
Also, $O_{i}^{(\nu)}$ is the \textit{number} of occurrence of character $\nu\in\Sigma$ in the $i$th string. For example, assume $\Sigma = \{ \texttt{A}, \texttt{B}, \texttt{C} \}$, and $S = \{ `\texttt{BCABAABC}', `\texttt{CAACBBAA}' \}$.
Here, the alphabet size is equal to $3$, $m = 3$. The number of strings, or $N$, is $2$. The first string is $s_1=`\texttt{BCABAABC}'$, the second string is $s_2=`\texttt{CAACBBAA}'$, the  remaining of $s_1$ after removing `$\texttt{B}$' is  $r_1^{\texttt{B}}=`\texttt{CABAABC}'$, while removing `$\texttt{B}$' from $s_2$ leads to $r_2^{\texttt{B}}=`\texttt{BAA}'$, and $R^{\texttt{B}}$ is $\{r_{1}^{B},r_{2}^{B}\}$. The number of occurrence of character `$\texttt{B}$' in $r_1^{\texttt{B}}$ is 2, or $O^{(\texttt{B})}_{1}$ = 2, and the  number of occurrence of character `$\texttt{B}$' in $r_2^{\texttt{B}}$ is 1, or $O^{(\texttt{B})}_{1}$ = 1.
Also, the following functions are upper bound ($ub$), mean ($\mu$), and variance ($var$), respectively.
\begin{equation*}
    ub(S) = \sum_{j=1}^{m} min \{ O_{i}^{(\sigma_{j})} \mid i = 1, \cdots, N  \},
\end{equation*}
\begin{equation*}
    \mu(R^{\nu}) = \frac{1}{N}\sum_{i=1}^{N} |r_{i}^{\nu}|,
\end{equation*}
\begin{equation*}
    var(R^{\nu}) = \frac{\sum_{i=1}^{N} {\left(|r_{i}^{\nu}| - \mu(R^{\nu}) \right)}^2 }{N - 1}.
\end{equation*}
These functions help introduce our proposed method, $GCoV$, in section~\ref{method}. The upper bound function $ub$ returns the highest possible length of the LCS. Each unique alphabet is counted among all strings in this function, and its minimum number is returned. This procedure is repeated for all the special characters in the alphabet, and the minimums are added together. The summation result is the upper bound. The $\mu$ and $var$ functions, as their symbols convey, are the mean and the variance of the strings' lengths, respectively.

\subsection{Beam search}
\begin{figure}
  \centering
  \includegraphics[width=\textwidth]{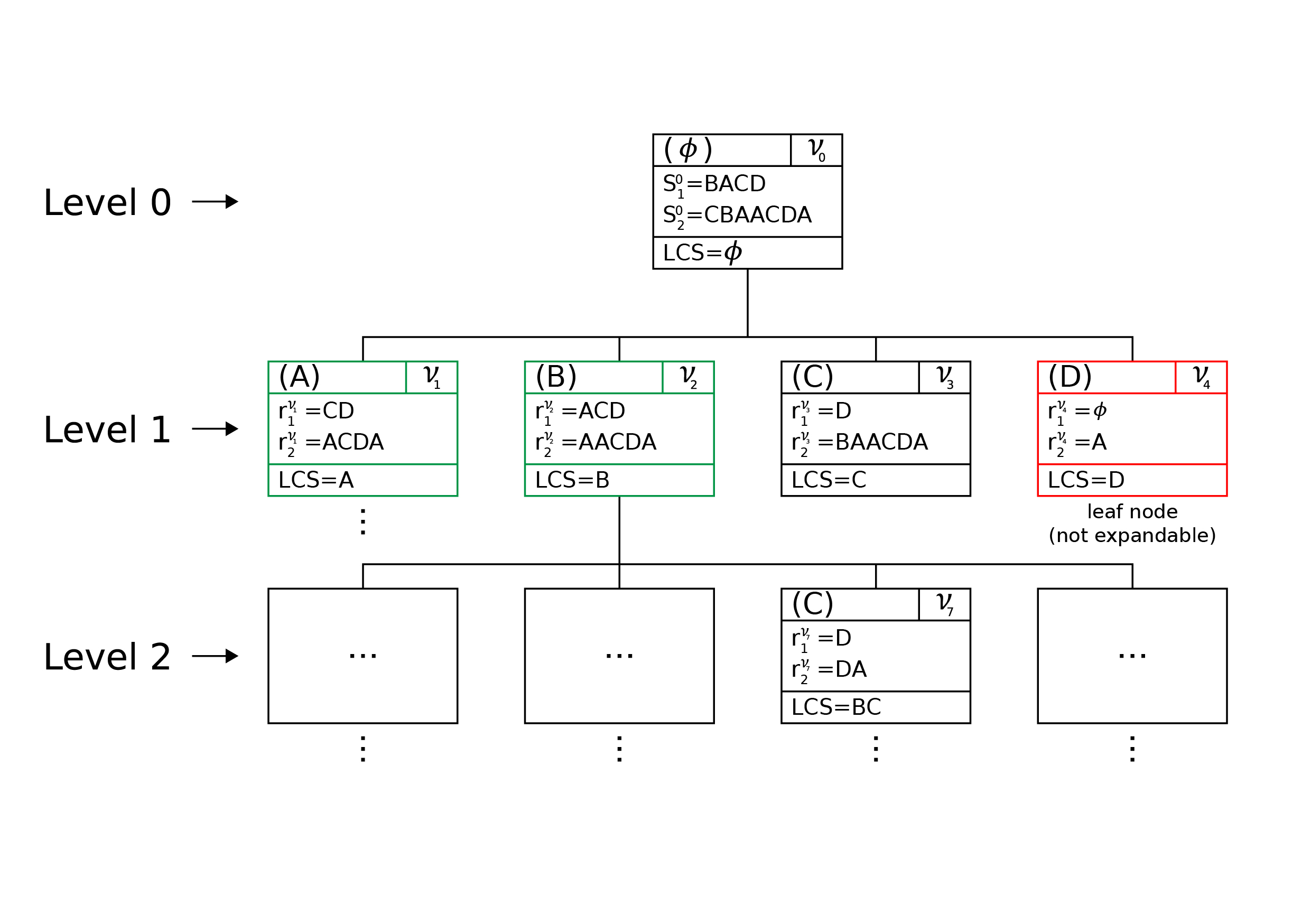}
   \caption{An example of beam search schema, where $\beta$ is 2. The root is called $\phi$ and contains the original set of strings. In the next level, each node shows a unique character from the alphabet set $\{A, B, C, D\}$. As an example, node $\nu_2$ shows the selection of the left-most $B$ as the next character of the LCS. Then, the mentioned $B$ and all its prior characters are removed from all strings and leads to $r_1^{\nu_2}=ACD$ and $r_2^{\nu_2}=AACDA$. This selection of $B$ causes the first letter of LCS becomes $B$, as shown at the bottom of the same node. By going from node $\nu_2$ to the node $\nu_7$, the character $C$ is picked as the next character of the LCS. This selection makes the LCS until that node equal to $BC$. In addition, the node $\nu_4$ is a leaf node. Because, choosing the $D$ leads to nullifying $r_1^{\nu_4}$. Thus, node $\nu_4$ becomes a leaf, and the longest possible common subsequence from that branch is equal to $D$, which is obviously less optimal than the branch of $\phi$-$\nu_2$-$\nu_7$. Suppose we set the score of each node equal to the best minimum of the strings. Thus, the nodes $\nu_1$ and $\nu_2$, shown in green, will be chosen as the ``best nodes'' for expansion in the next step.}
  \label{fig:bs}
\end{figure}

Beam search is a version of the breadth first search having a constraint on the number of nodes in each tree level. It has a parameter $\beta$, which controls completeness and greediness. In other words, the $\beta$ specifies the limit on the number of nodes that are allowed to be expanded in the next step of the algorithm. Thus, the problem is formalized in a tree search form in which edges are a selection of a unique alphabet character as the part of the LCS. Therefore each node has $|\Sigma|$ children. The children nodes contain the remaining strings ($R^{\nu}$),  after removing the corresponding letter. In each level, $\beta$ best nodes among all children are chosen based on a scoring function. 
For example, suppose $\beta = 2$, as Fig.~\ref{fig:bs} shows at level 1, we choose the letters $\nu_{1}$ and $\nu_{2}$ due to having higher scores than others. These scores can be acquired using different heuristics, e.g., minimum length of the remaining strings, upper bound, or probabilistic scores~\cite{blum09,mousavi12improved,djukanovic19}. 
 \newline
 Algorithm~\ref{alg:1} shows the steps of the beam search algorithm for the LCS problem. Beam search starts with an empty node and constructs the solutions until no expandable node exists. It consists of three major phases. In the first phase, the candidate nodes in the $BestNodes$ list are expanded by utilizing the $\Sigma$ members, and if they are feasible, they are added to the expanded nodes list ($ChildNodes$). In the second phase, a heuristic function is employed to evaluate nodes in $ChildNodes$  and assigns scores to them. In the last phase, $\beta$ nodes of $ChildNodes$ are selected according to their scores and put in the $BestNodes$ list to be expanded for the next step of the algorithm.
 
\begin{algorithm}
\caption{Beam Search (BS) for LCS}\label{alg:1}
\begin{algorithmic}
\Require $S$ (Set of Strings), $\Sigma$ (Alphabet), $\beta$ (Beam width), and $HF$ (Heuristic Function)

\State //Initialization: $BestNodes = \{\phi\}$

\While{$BestNodes$ is not empty}
    \State \#Expansion phase
    \For{each $b_{i}$ in $BestNodes$}
        \State $ChildNodes$ = $ChildNodes$ $\cup$ \texttt{Successor}($b_i, \Sigma$)
    \EndFor
    \State \#Score evaluation phase
    \For{each $c_i \in ChildNodes$}
        \State $Score_{i} = \texttt{HF}$($c_i$)
    \EndFor
    \State $BestNodes = \{ \}$
    
    \State \#Best nodes selection phase
    \State $BestNodes = \texttt{BestNodesSelection}$($ChildNodes, Score, \beta$)
    \State $lcs = \texttt{LongestSolution}$($BestNodes$)
\EndWhile \\
\Return $lcs$
\end{algorithmic}
\end{algorithm}

\section{Related work}\label{rewo}
This section presents the main and most effective studies done in the field of the longest common subsequence. We should note that there are plenty of methods for solving the LCS problem, but few of these methods can be used in the real world. Among all approaches, as the results show and other experts mentioned~\cite{blum09}, beam search is suitable in performance as well as run-time. So, we describe the most promising methods based on beam search. For the first time, it was Blum et al.~\cite{blum09} who proposed the beam search to solve the LCS problem. Their heuristic for score computation in each node was simply the length of the minimum remaining string, which showed the power of utilizing beam search for solving the LCS. 

Later, Mousavi and Tabataba~\cite{mousavi12improved} put the foundation of their work on the proposal of Blum et al.~\cite{blum09}. Mousavi and Tabataba assumed that the strings in $S$ are independent of each other.
Regarding this assumption, they defined a probabilistic recursive function $p(k,n)$, where it is the probability of having the LCS with length $k$ in a string with length $n$. They computed this probability for each string in the set of strings $S$ and then multiplied those probabilities. Since then, this probability table has been the building block of the LCS problem. Tabataba and Mousavi~\cite{tabataba12hyper}, in another work, proposed a hyper-heuristic method for solving the LCS problem. To the best of our knowledge, this hyper-heuristic is the only hyper-heuristic for solving the LCS problem. Also, they introduced an improved version of the Blum et al.~\cite{blum09} heuristic, which is a nonlinear combination of the minimum length of remaining strings and the length of all remaining strings. Their hyper-heuristic applied both heuristics on the dataset with a small $\beta$, and the heuristic with the longest solution was chosen for the final evaluation with the bigger $\beta$. 

In 2019, Djukanovic et al.~\cite{djukanovic19} introduced an expected length heuristic ($BS$-$Ex$) that uses the probability table introduced by~\cite{mousavi12improved}. Because of negligible values of probabilities of their heuristic, they had to utilize approximations in their proposals. As the authors mentioned in their paper, in addition to the function approximation, they used a divide and conquer method to improve the run-time of their algorithm. It is worth mentioning that divide and conquer, in addition to affecting run-time, enhances the quality of solutions.
Moreover, their approximation cannot work properly due to that they have not had a closed-form equation of $p(k,n)$. In this paper, as our main contribution, we introduce a closed equation of the probability function $p(k,n)$ in a theorem~\ref{closedEq1} section~\ref{equat}.  
Their heuristic obtained the best results for ``uncorrelated'' and ``almost uncorrelated'' strings. But, the authors' proposal has a low performance for correlated strings. 

Nikolic et al.~\cite{Nikolic21} presented the $GMPSUM$ heuristic function, which consists of two different functions as follows. The first one ($GM$) is based on the computation of the geometric mean over the geometric variance of the remaining lengths of strings. The other one ($PSUM$) considered sum of probabilities $p(k,n)$ for different values of $k$ and $n$. In other words, the latter is a summation of all possible probabilities. Then, they use a convex combination of two score functions as a heuristic function. The authors reported the best possible convex coefficients for each dataset. The computation of coefficients needs a hard fine-tuning for each dataset. Selecting the suitable coefficient problem makes their method inapplicable in real problems. The results, as the authors showed, are proper for correlated strings. In any event, their method needs a considerable amount of computations which causes its inefficiency in execution. In addition, their method leads to overflow error during computing geometric mean with $N\geq~150$ and $|\Sigma|=4$, which produces a value greater than $550^{150}$ which is an incredibly huge number.

\section{Theorems on equational formulation of tabular probability computation}\label{equat}
In this section, we introduce the equational form of the randomized tabular probability introduced by Mousavi and Tabataba~\cite{mousavi12improved}. They showed that the probability of having a common subsequence of length $k$ in a string with length $n$ is equal to Eq.~\ref{eq:taba}.
 \begin{equation}
     p(k, n) = 
     \begin{cases}
     1 & \text{k = 0}\\
     0 & \text{k $>$ n}\\
     \alpha \times p(k - 1, n - 1) + \beta \times p(k, n - 1) & else,
     \end{cases}
     \label{eq:taba}
 \end{equation}
where $\alpha = \frac{1}{|\Sigma|}, \beta = \frac{|\Sigma| - 1}{|\Sigma|}, \alpha + \beta = 1$. 
So by defining the initial values and utilizing recursive calculations, each value of $p(k, n)$ can be acquired. Then, they proposed a tabular form to compute this probability and called it dynamic programming.
In other words, this probability had been recursively computed and saved in a table. From Mousavi and Tabataba's work, most researchers on this subject have put their effort into improving the performance of finding LCS using Eq.~\ref{eq:taba}. One of the best results has been reported by Djukanovic et al.\cite{djukanovic19}. The authors tried to approximate the score and reached a higher performance, if not the best. The probability Eq.~\ref{eq:taba} has been very promising for the LCS problem. However, due to not having the closed equational form of the probability computation, their estimation does not have a closed, well-formed equational basis. Additionally, if there exists a method of direct computation of Eq.~\ref{eq:taba}, it is not necessary to save the tabular form when there is a memory limit problem. Our first contribution is to find a closed-form of Eq.~\ref{eq:taba}. The below theorem shows this. \newline
\textbf{Theorem 1}- By having $\alpha = \frac{1}{|\Sigma|}, \beta = \frac{|\Sigma| - 1}{|\Sigma|}, \alpha + \beta = 1$,  the probability of finding a subsequence of length $k$ in a string with length $n$, Eq.~\ref{eq:taba}, is equal to 
    
\begin{equation}
    p(k, n) = 1 - \beta^{n-k+1}  \left[ \sum_{i = 0}^{k - 1} \alpha^i {n-k+i \choose i} \right].
\label{closedEq1}
\end{equation}

Proof of the theorem 1 comes in~\ref{appendix1}. Theorem 1 can be stated in other forms. \newline
\textbf{Corollary 1}- The probability of finding a subsequence with length $k$ in a string with length $n$, $p(k, n)$, can be stated in two different forms of
\begin{center}
\begin{enumerate}[I]

    \item) 
    \begin{equation*}
    p(k, n) = 1 - \beta^{n-k+1}  \left[ \sum_{i = 0}^{k - 1} \alpha^i {n-k+i \choose i} \right]  = 1 - \beta^{n-k+1}  \left[ \sum_{i = 0}^{k - 1} \alpha^i {n-k+i \choose n - k} \right]; 
    \end{equation*}
    
    \item) 
    \begin{equation*}
    p(k, n) = 1 - \beta^{n-k+1} - \beta^{n-k+1} \left[ \sum_{i = 0}^{k-1} \prod_{j=1}^{i} \alpha \frac{n-k+j}{j} \right].
    \end{equation*}    
\end{enumerate}
\end{center}
Proof of corollary~1 is simply derived from the properties of the combination. 
Theorem~1 and its direct Corollary~1 give an equivalent closed-form of Eq.~\ref{eq:taba}. The Eq.~\ref{closedEq1} and its siblings not only relieve the need for recursive computation but also open a way for better analysis and approximation of the LCS. We have used the results of manipulating the closed-form Eq.~\ref{eq:taba} to get better results in section~\ref{method}. Thus, we define $q(k,n)$ as follows.  
\begin{equation}
    \label{eq:qkn}
    q(k, n) = \frac{1 - p(k, n)}{\beta^{n-k+1}} = \left[ \sum_{i = 0}^{k - 1} \alpha^i {n-k+i \choose i} \right].
\end{equation}

It is necessary to mention that we have found another interesting closed-form of the Eq.~\ref{closedEq1} during our observations and analysis. The probability $p(k, n)$ is nothing but a weighted sum of $Beta$ density functions. The below theorem states this claim.
 
\textbf{Theorem 2}- The probability of finding a subsequence with length $k$ in a string with length $n$ can be stated as the sum of $Beta$ density functions, 
\begin{equation}
p(k, n) = 1 - \beta^{n - k + 1} - \alpha \beta \left[ \sum_{i = 1}^{k - 1} \frac{1}{i} Beta(\beta, n-k, i) \right].
\end{equation}
A proof is provided for theorem 2 in \ref{appndix2}

\section{Proposed methods and new analysis}\label{method}
\begin{figure}[H]
  \centering
  \includegraphics[width=\textwidth]{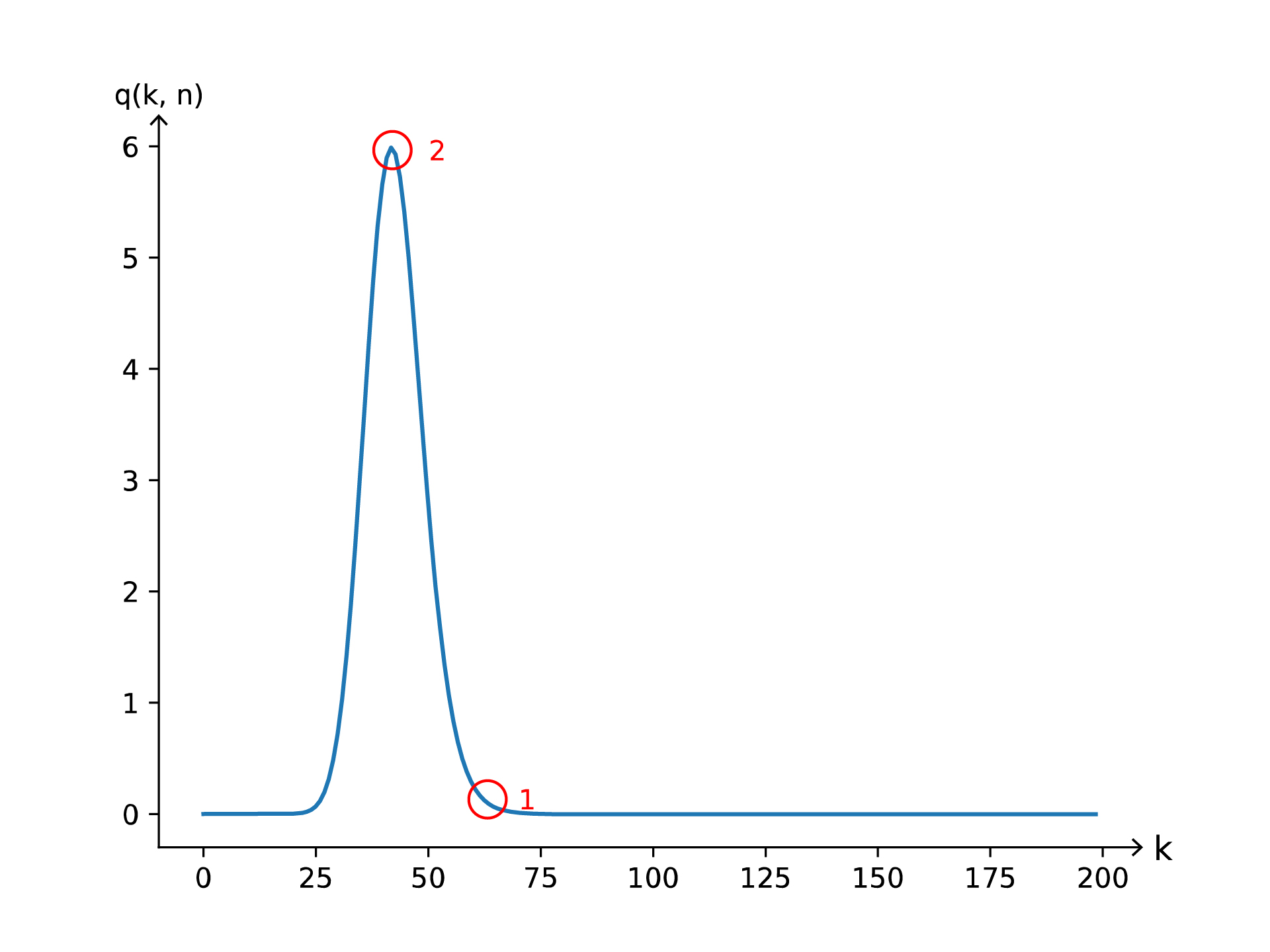}
   \caption{Plot of $q(k, n)$, where $n = 200$, $k \in \mathbb{N}$, and $0\leq k \leq n$. Points are used in this paper as $k_{analytic}$ are shown with number 1 for uncorrelated strings and number 2 for correlated strings.}
  \label{fig:qkn}
\end{figure}

\begin{figure}
\begin{subfigure}{.5\textwidth}
  \centering
  \includegraphics[width=\textwidth]{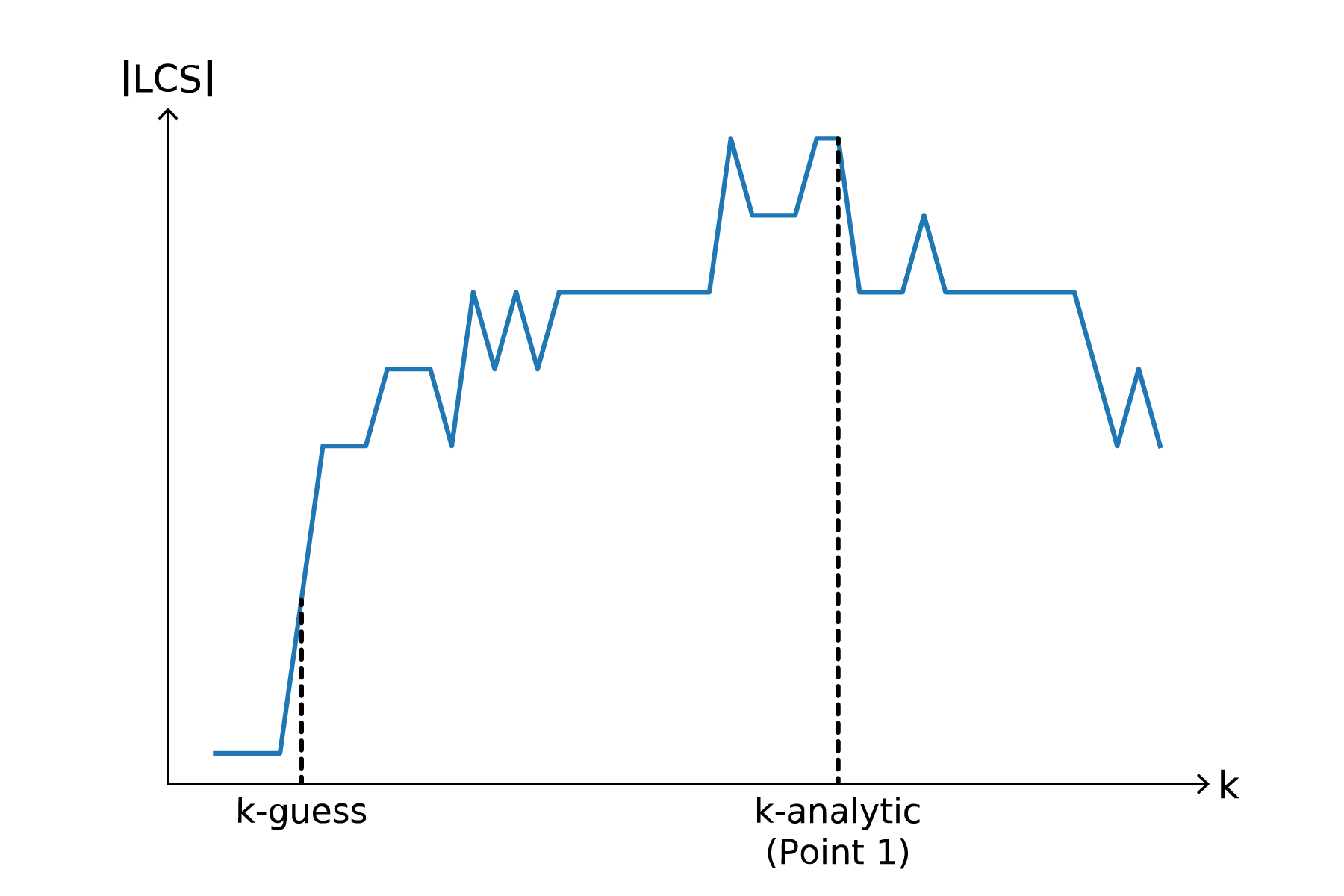}
  \caption{ }
  \label{fig:difka}
\end{subfigure}%
\begin{subfigure}{.5\textwidth}
  \centering
  \includegraphics[width=0.88\textwidth]{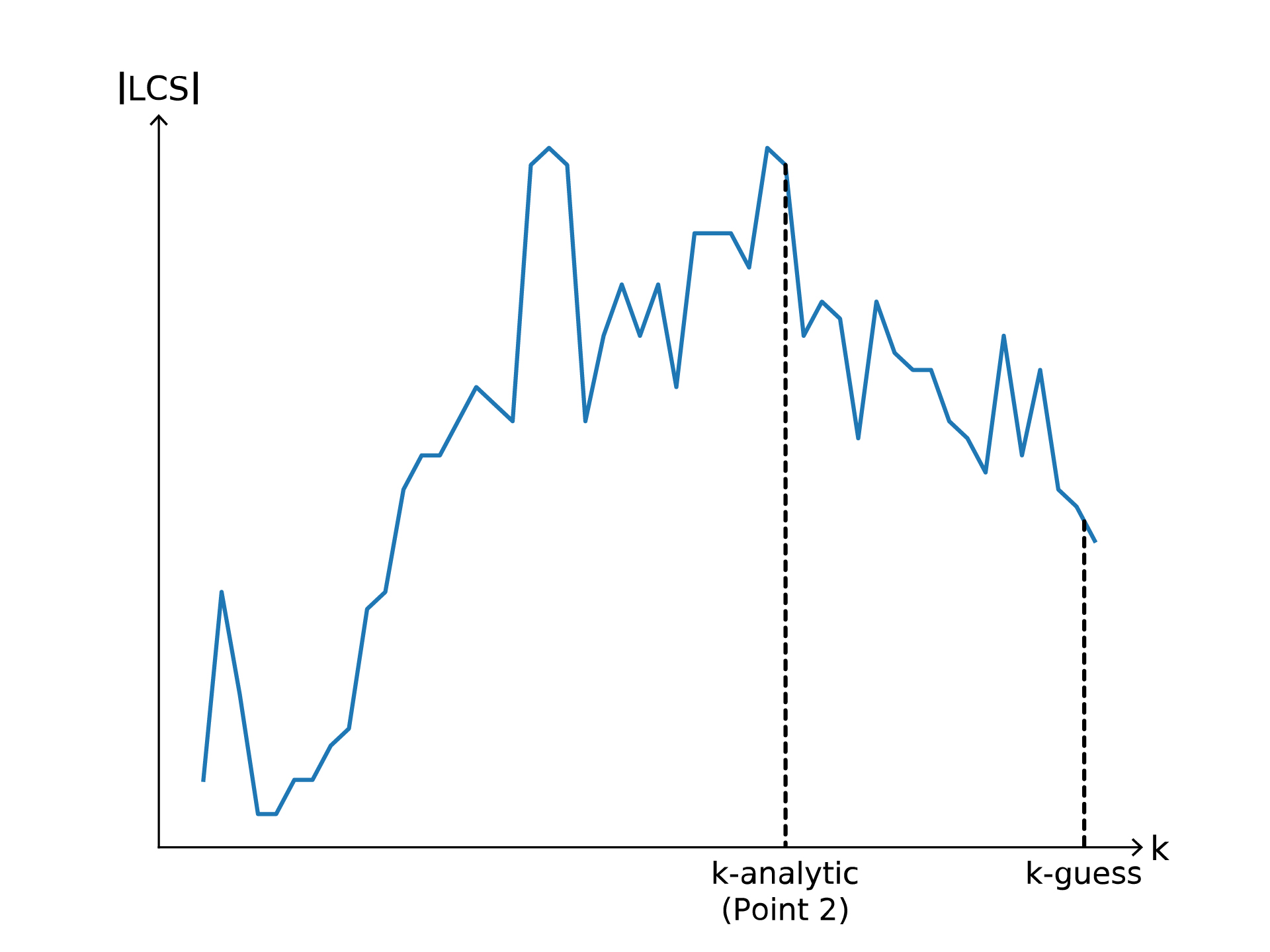}
  \caption{}
  \label{fig:difkb}
\end{subfigure}
\caption{Lengths of LCS based on the allocation of different values of $k$. a) Results for an uncorrelated dataset. The length of LCS  becomes shorter for the points that go farther from the value of the $k_{analytic}$ (point~1). b) Results for an uncorrelated dataset. The length of LCS  becomes shorter for the points that go farther from the value of the $k_{analytic}$ (point~2)}
\label{fig:difk}
\end{figure}

This section deals with the power that the closed-form equation provides. Also, we introduce the $GCoV$ heuristic function. At last, we use another method that takes advantage of the closed-form analytic method and $GCoV$. We use beam search as our search strategy. Mousavi and Tabataba~\cite{mousavi12improved} suggested using a probability score function for each node $\nu$. To compute the score, they defined a recursive function $p(k, |r_{i}^{\nu}|)$ which is the probability of having a subsequence with length $k$ in the remaining of each string by choosing node $\nu$. By having $p(k, |r_{i}^{\nu}|)$ for all strings, they multiplied these probabilities as the scoring function, or
\begin{equation}
    h(\nu) = \prod_{i = 1}^{N} p(k, |r^{(\nu)}_i|).
    \label{eq:6}
\end{equation}
In the above formula, $k$ estimates the length of the LCS of the remained postfix of the strings, and specifying $k$ is not a trivial task. Mousavi and Tabataba~\cite{mousavi12improved} assumed that as the size of the alphabet set is larger, the length of the LCS will be shorter. Moreover, the chance for a longer LCS is increased by having the longer remaining postfix of strings. Using these two assumptions, they guessed $k=\frac{min \{|R_i^{\nu}|\} }{|\Sigma|}$, where $i=\{1,...,|ChildNodes|\}, \nu \in ChildNodes$. It is notable that Jiang and Li~\cite{jiang95} estimated the expected length of the LCS problem for $N$ uncorrelated string of length $N$ to almost $\frac{N}{|\Sigma|}$. We call obtained $k$ by Mousavi and Tabataba as $k_{guess}$ in our paper. Due to the lack of a closed-form equation for $p(k, n)$, Mousavi and Tabataba could not analyze the value of $k$. Instead, we use the closed-form equation of $p(k,n)$ to find an analytic way of estimating $k$. It can be acquired by checking the plot of  $p(k,n)$ or its equivalent,~$q(k,n)$ from Eq.~\ref{eq:qkn}. Fig.~\ref{fig:qkn} shows the plot of~$q(k,n)$. The two points of point~1 and point~2 show interesting behaviors as $k$ for different types of datasets.
First of all, we choose values of $k$ between ``point~1'' and ``point~2'' and call our obtained $k$ to $k_{analytic}$. The analytic investigations show that choosing $k$ almost equal to point~1 leads to better results and consequently improves the performance of finding near-optimal LCS of \textit{uncorrelated} and \textit{almost uncorrelated} strings. Surprisingly, as $k$ goes almost near to point 2, we obtain better results for \textit{correlated} strings. We plot the results of $k_{guess}$ and $k_{analytic}$ for both types of uncorrelated vs correlated datasets in Fig.~\ref{fig:difk}. As Fig.~\ref{fig:difka} shows, a closer value of $k$ to point~1 returns the best results among all possible results for an uncorrelated dataset. On the other hand, as Fig.~\ref{fig:difkb} shows, a closer value of $k$ to point~2 returns the best results among all possible results for a correlated dataset. These two points are two extreme points of the spectrum of choosing proper and analytical $k$. In other words, for different types of datasets, from uncorrelated to the correlated datasets, $k$ can be acquired by moving on the graph in Fig.~\ref{fig:qkn} from point~2 to point~1.   All of these results are the consequences of finding the closed-form of computing $p(k,n)$. 

It is possible to reach an estimation of $k_{analytic}$. In this paper, we  apply  simple regression methods and Eq.s~\ref{eq:uniformeq}, and~\ref{eq:nonuniformeq} are obtained as follows. 

\begin{equation}
    k_{analytic} = \frac{max\{|R_i^{\nu}|\} * (a - \left( b \times {\ln N} \right) )}{|\Sigma|} 
    \label{eq:uniformeq}
\end{equation}
\begin{equation}
    k_{analytic} = \frac{min\{|R_i^{\nu}|\} - c}{|\Sigma|},
    \label{eq:nonuniformeq}
\end{equation}
where $a = 1.8233, b = 0.1588$, $c = 31$, and $i=\{1,...,|ChildNodes|\}, \nu \in ChildNodes$. Eq.~\ref{eq:uniformeq} is for uncorrelated strings. It depends on $N$ as well. On the other hand, the Eq.~\ref{eq:nonuniformeq} is for correlated strings and it is independent from the value of $N$. It is important to note that the main assumption behind the heuristic function~\ref{eq:6} is the independence of sequences in $S$.
However, here  $k_{analytic}$ can reach the best solution among all approaches for correlated strings. The last thing to mention, we set $k_{analytic}$ to 1, when its value is equal or smaller than $0$ in the both Eq.s~\ref{eq:uniformeq} and~\ref{eq:nonuniformeq}. \newline
Another contribution of this paper is a heuristic function based on the Coefficient of Variation$\footnote{Coefficient of Variation (CoV) is a standardized measure of the dispersion of a probability distribution or frequency distribution~\cite{cov98}}$. We call this heuristic ``Generalized Coefficient of Variation'',   or $GCoV$. A node $\nu$ with a higher mean, lower variance, and a higher amount of upper bound shows better behavior to be explored in the next step. Thus, we can formalize the well-behaving as a heuristic as follows.
\begin{equation}
    GCoV(\nu) = \frac{ \left[\mu(R^{\nu})\right]^{2}}{\left[var(R^{\nu})\right]^{\gamma}} \times \sqrt{ub(R^{\nu})},
    \label{eq:cov}
\end{equation}
where $\mu(R^{\nu})$ is the mean , $var(R^{\nu})$ is the variance, $ub(R^{\nu})$ is the upper bound function. These functions are defined in section~\ref{req}. The $\gamma$ that we have used in Eq.~\ref{eq:cov} changes for different $N$'s. As $N$ grows, the utility of the variance increases. The value of $\gamma$ is obtained using trial and error. To obtain the general value of  $\gamma$, a linear regression was applied, and the following function is obtained $\gamma = (0.0036 \times N) - 0.0161$. Our observations show that the methods of $k_{analytic}$ and $GCoV$ are almost complementary. So, we have decided to use the hyper-heuristic, introduced in~\cite{tabataba12hyper}, for these two heuristic functions. In the first phase, this hyper-heuristic runs both methods but with a smaller size of beam width, $\beta_h$. Then, among the methods, the one with the better response is chosen to perform the complete version of the beam search with actual $\beta$. The~Algorithm~\ref{alg:2} shows the hyper-heuristic steps.

\begin{algorithm}
\caption{Hyper-Heuristic algorithm for the LCS problem}\label{alg:2}
\begin{algorithmic}
\Require $S$, $\Sigma$ , $\beta$, $\beta_h$ (small beam width), $HF_1$, and $HF_2$, 
\State $LCS_{short1} = BS(S, \Sigma, \beta_{h},HF_1)$
\State $LCS_{short2} = BS(S, \Sigma, \beta_{h},HF_2)$
\If{$|LCS_{short1}|\geq |LCS_{short2}|$}
\State \Return $BS(S, \Sigma, \beta,HF_1)$
\Else
\State \Return $BS(S, \Sigma, \beta,HF_2)$
\EndIf
\end{algorithmic}
\end{algorithm}

\section{Experimental results}\label{results}
In this section, we report the performance of our proposed methods with the best approaches in the literature on short-run methods. ``Short-run'' methods do not require excessive amounts of memory and time. The long-run algorithms are complete and anytime algorithms, e.g., $A^{*}$. The latter algorithms improve their results by having a higher amount of memory as well as time~\cite{Zilberstein96}.
We ran our experiments on a Windows 10 platform with a specification of 6 GB of RAM and CPU i7-4702MQ. We used benchmark datasets of the LCS problem, i.e., ACO-Rat, ACO-Virus, and ACO-Random from~\cite{shyu009}. In addition, we used BB dataset from~\cite{blum07}. The last dataset that we have utilized is $SARS$-$CoV$-$2$ genomes. We implemented our proposed methods, i.e., $k_{analytic}$, $GCoV$, and the Hyper-heuristic (HH) of those methods. To evaluate our results, we have implemented the best short-run methods proposed in the LCS literature, including $k_{guess}$~\cite{mousavi12improved}, $BS$-$Ex$\cite{djukanovic19}, and $GMPSUM$\cite{Nikolic21}. All methods here are implemented using the $Python$ version 3.8 programming language. The beam width~($\beta$) and small beam width~($\beta_{h}$) are set to 200 and 60, respectively, for all of the runs on all methods. To show the impact of choosing $k$ properly, first, we report the comparison between $k_{guess}$ and $k_{analytic}$~\cite{mousavi12improved}. Then, we compare the proposed methods with the best available methods of the LCS.

\subsection{ Comparison of $k_{guess}$ \textit{vs.} $k_{analytic}$} 
Here, we compare $k_{guess}$ and $k_{analytic}$ to show the impact of the closed analytical form of $p(k,n)$ computation. Results of Tables~\ref{tb:t1},~\ref{tb:t2},~\ref{tb:t3}, and~\ref{tb:t4} show that choosing $k$ according to our analysis outperform the $k$ obtained by Mousavi and Tabataba~\cite{mousavi12improved}. However, in terms of improvement percentage  (IP) on average for each dataset, the IP in Rat dataset is $2.31\%$, the Virus dataset is $0.52\%$, the Random dataset is $1.06\%$, and BB dataset is $1.66\%$. The comparison of $k_{analytic}$ versus  $k_{guess}$ comes for the Rat dataset in Table~\ref{tb:t1}. More results on other datasets are in Tables~\ref{tb:t2},~\ref{tb:t3}, and~\ref{tb:t4} in ~\ref{appndix3}.

\begin{table}[H]
\centering
\caption{Comparison of $k_{guess}$ and $k_{analytic}$ on Rat dataset}
\label{tb:t1}
\begin{tabular}{lllll}
\hline
\textbar{}$\Sigma$\textbar{} & N   & l             & $k_{guess}$\cite{mousavi12improved} & $k_{analytic}$
\\
\hline
4                         & 10  & 600         &198  & \textbf{202}      \\
4                         & 15  & 600       &183  & \textbf{184}      \\
4                         & 20  & 600             & 167  & \textbf{172}      \\
4                         & 25  & 600            & 166 & \textbf{169}      \\
4                         & 40  & 600          & 142 & \textbf{151}      \\
4                         & 60  & 600             & 147 & \textbf{152}      \\
4                         & 80  & 600             & 137& \textbf{139}      \\
4                         & 100 & 600             & 134  &\textbf{135}     \\
4                         & 150 & 600            & 124  & \textbf{129}      \\
4                         & 200 & 600             & 121  & \textbf{122}      \\ \hline
20                        & 10  & 600             & 68  & \textbf{70}       \\
20                        & 15  & 600            & 61  & \textbf{62}       \\
20                        & 20  & 600            & 53  & \textbf{54}       \\
20                        & 25  & 600            & 50  & \textbf{51}       \\
20                        & 40  & 600            & 49 & 49       \\
20                        & 60  & 600            & 46 & 46       \\
20                        & 80  & 600            & 43  & 43       \\
20                        & 100 & 600            & 38 & \textbf{39}       \\
20                        & 150 & 600            & 35 & \textbf{37}       \\
20                        & 200 & 600            & 32  & \textbf{34}   
\\ \hline
Average                   & \multicolumn{2}{l}{} & 99.7 &\textbf{102}  \\
\hline
\end{tabular}
\end{table}

\subsection{Comparison of proposed methods with the state-of-the-art methods}
In addition to showing the performance of $k_{analytic}$ over $k_{guess}$, the theorems 1 and 2 and their direct analysis help to create new methods. These methods take advantage of the closed-form equations and the numerical behavior of the remaining strings, $R^{\nu}$. The methods are $k_{analytic}$ as well $GCoV$, both introduced in section~\ref{method}. Moreover, we use a hyper-heuristic that makes use of both mentioned methods. To the best of our knowledge, the methods $BS$-$Ex$\cite{djukanovic19} and  $GMPSUM$\cite{Nikolic21} presented the best results on benchmark datasets. Thus, we compare our results with them in Tables~\ref{tb:rat},~\ref{tb:virus},~\ref{tb:rnd},~\ref{tb:bb}, and~\ref{tb:sars_cov_2}. The following Tables show that our proposed methods can obtain a better average solution over all benchmarks. 
\newline
The $k_{analytic}$ for uncorrelated datasets (Rat, Virus, Random) obeys the Eq.~\ref{eq:uniformeq}. As the results in Tables~\ref{tb:rat}, ~\ref{tb:virus}, and~\ref{tb:rnd}  show some values of $GMPSUM$\cite{Nikolic21} are left empty, especially for lengthier strings. It is impossible \textit{for us} to compute some values of $GMPSUM$. As mentioned in section~\ref{rewo}, the method produces huge numbers, which lead to overflow error in general PCs, i.e., it produces numbers equal to $550^{150}$ and above that. As the length of strings in $S$, or $N$, increases, the $GMPSUM$ produces larger numbers, and its computation becomes too complicated. To be more precise, the $GMPSUM$ was proposed as the combination of GM and PSUM. The GM section produces huge numbers and consequently makes the method inapplicable. On the other side, the PSUM section needs a considerable amount of time to prepare the results and thus leads to a lower efficiency for the whole method of $GMPSUM$. 

\begin{figure}[H]
\begin{subfigure}{.5\textwidth}
  \centering
  \includegraphics[width=\textwidth]{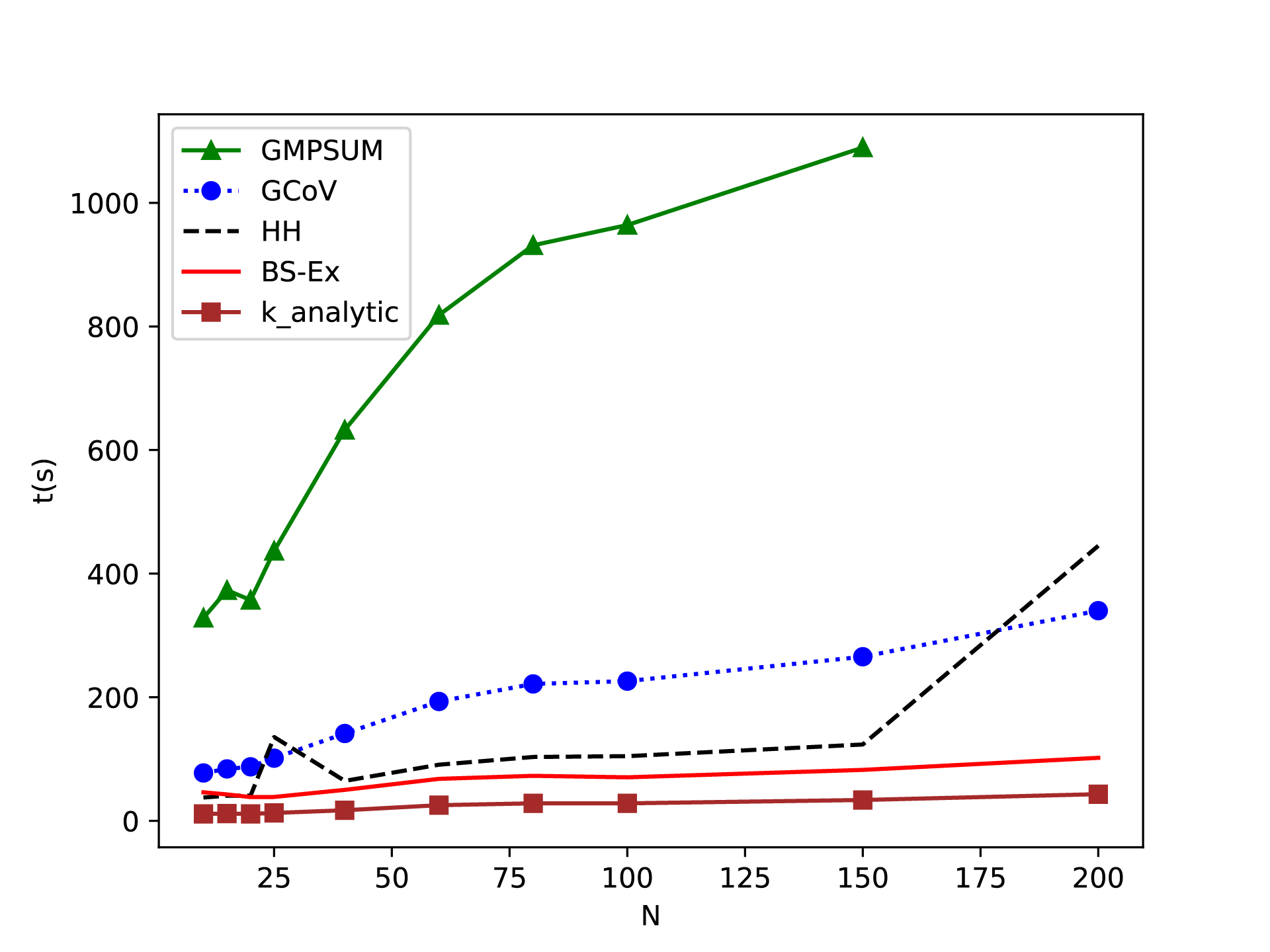}
  \caption{On Rat dataset }
  \label{fig:timerat}
\end{subfigure}%
\begin{subfigure}{.5\textwidth}
  \centering
  \includegraphics[width=\textwidth]{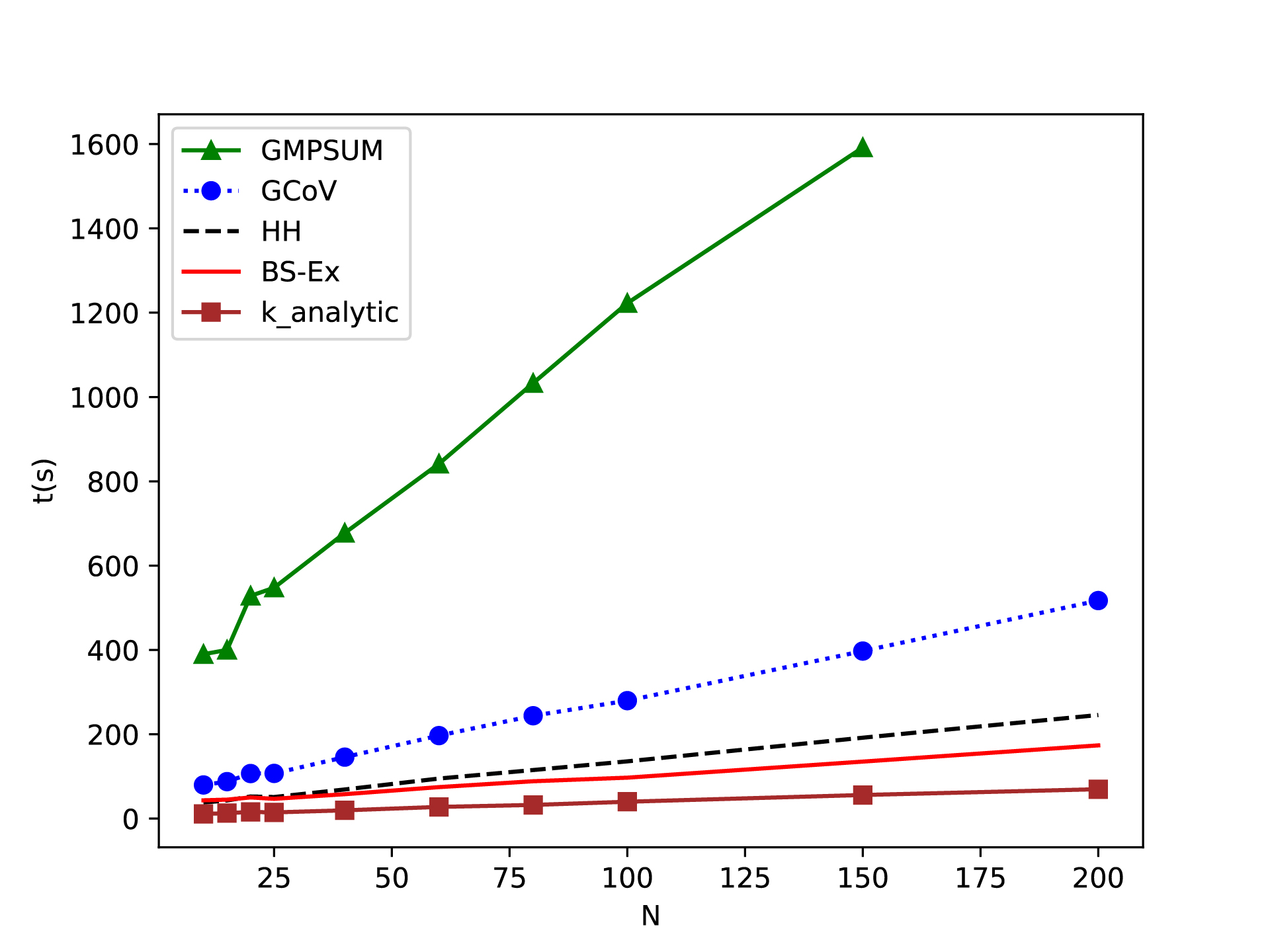}
  \caption{On Virus dataset }
  \label{fig:timevirus}
\end{subfigure}%

\begin{subfigure}{.5\textwidth}
  \centering
  \includegraphics[width=\textwidth]{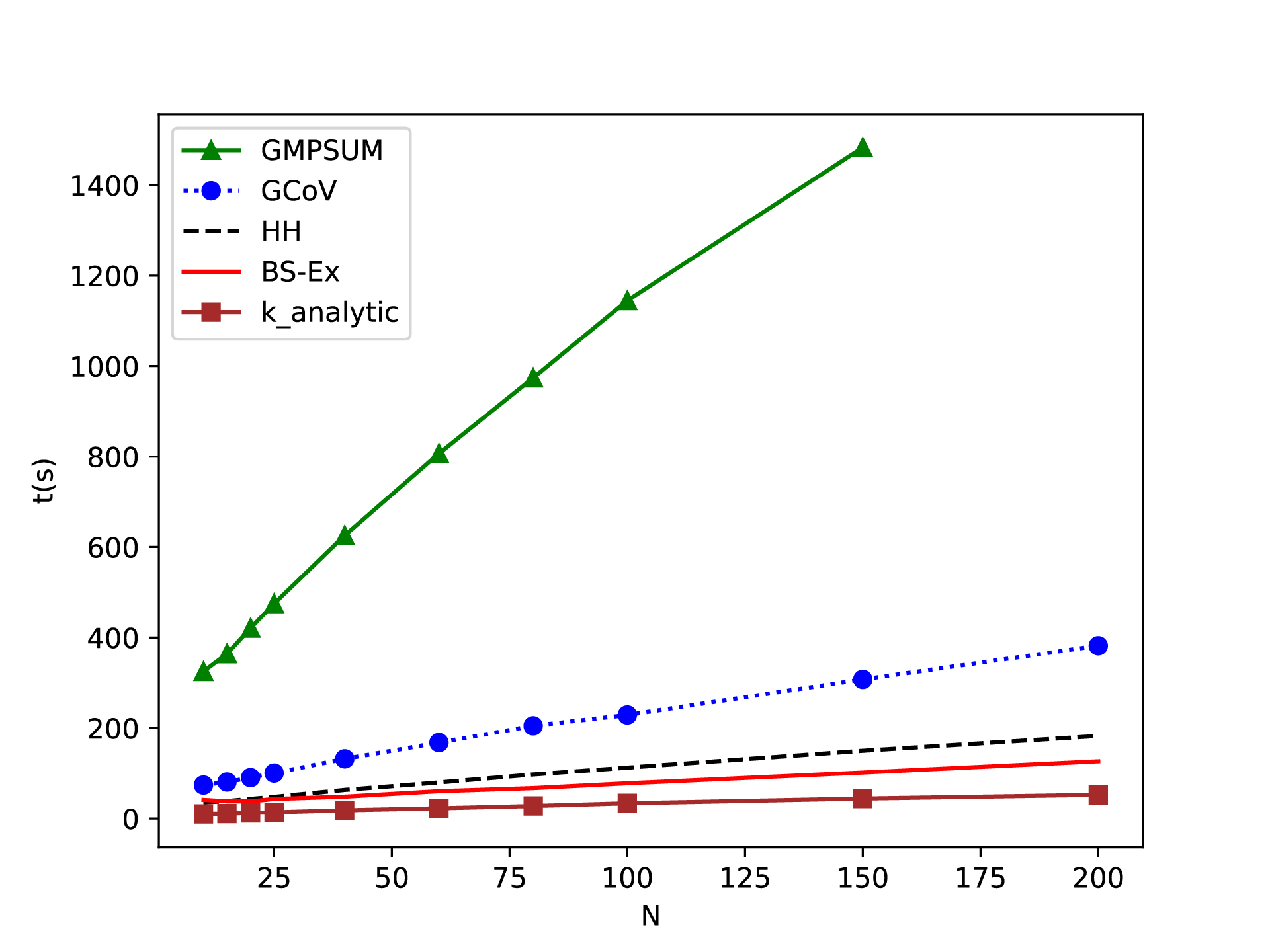}
  \caption{On Random dataset }
  \label{fig:timerandom}
\end{subfigure}%
\begin{subfigure}{.5\textwidth}
  \centering
  \includegraphics[width=\textwidth]{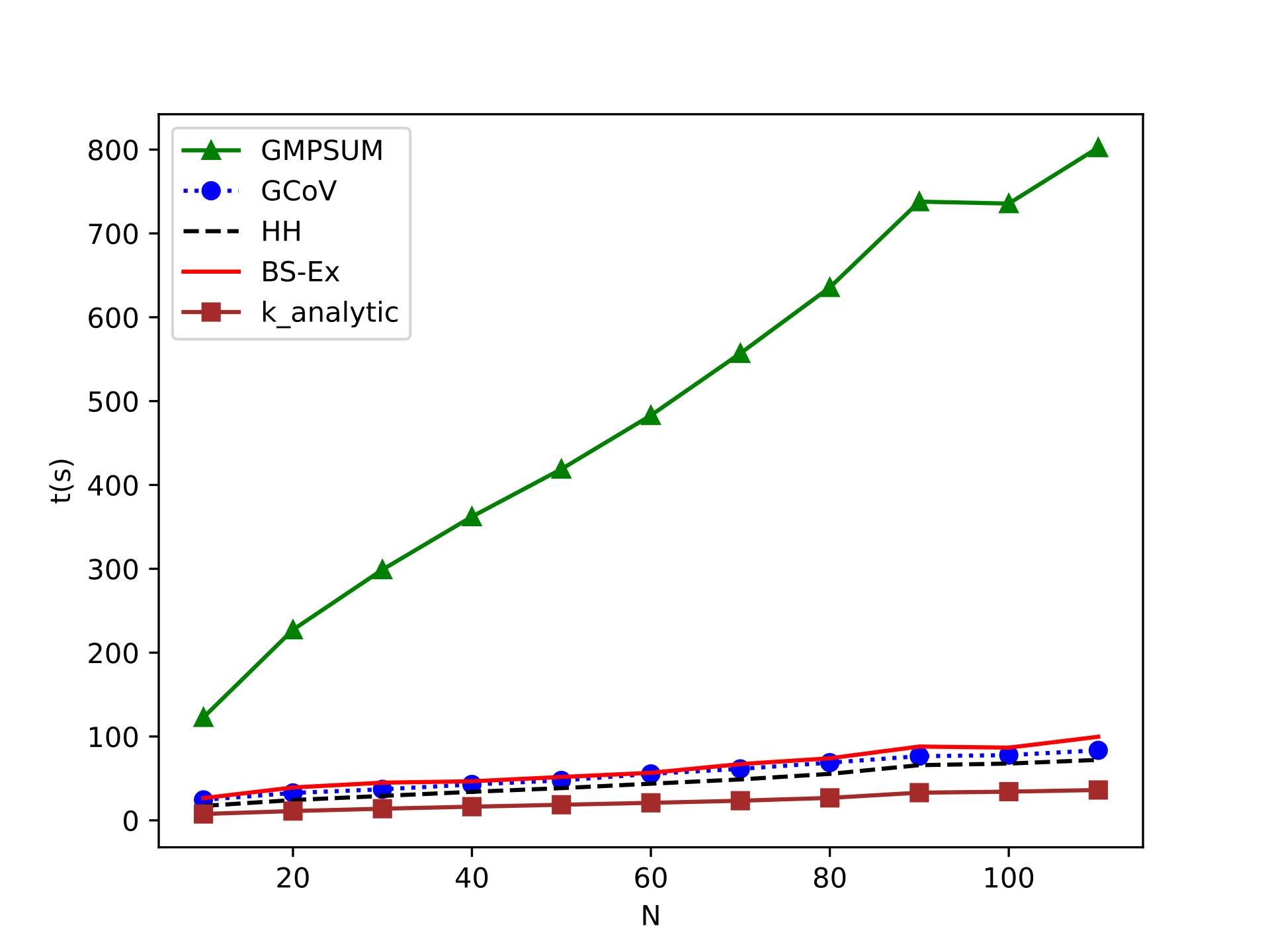}
  \caption{On SARS-CoV-2 dataset }
  \label{fig:timesars}
\end{subfigure}%
\caption{Comparison the running time of the LCS methods. The GMPSUM method for (a), (b), and (c) is not computable for $N=200$. The best result belongs to the $k_{analytic}$ and the worst belongs to $GMPSUM$.}
\label{fig:time}
\end{figure}

The $GM$ part in $GMPSUM$ and $GCoV$ can be considered from the same family. Both of them are a combination of mean, variance, and upper bound. However, the former uses a geometric interpretation of them, and the latter uses an arithmetic version of those statistics and factors. Thus, while the $GM$ needs more time and leads to huge numbers, the $GCoV$ consumes lower resources of time and space. Moreover, the results of $GMPSUM$ come from a brute-force search of coefficients, while $GCoV$ needs no tuning at all. Moreover, $GCoV$ is computable for any size of strings and any size of the alphabet. As we have shown in Fig.~\ref{fig:time}, In opposite to $GMPSUM$, the $GCoV$ is usable in any situation with a lower resource. It is good to mention that we did aim to put $GCoV$ instead of $GM$ in the $GMPSUM$. 
However, the size of the coefficients makes this combination inapplicable.

\begin{table}[H]
\centering
\caption{Comparison of state-of-the-art methods vs. our proposed methods on Rat dataset}
\label{tb:rat}
\begin{tabular}{llllllll}
\hline
\textbar{}$\Sigma$\textbar{} & N   & l & $BS$-$Ex$\cite{djukanovic19}  & $GMPSUM$\cite{Nikolic21} & $k_{analytic}$ & $GCoV$ &HH 
\\
\hline
4                         & 10  & 600            &201  &198 &\textbf{202}&198 & 202 \\
4                         & 15  & 600            &183  &\textbf{184}&\textbf{184}&182  & 184      \\
4                         & 20  & 600            &171  & \textbf{173}&172&169  & 172      \\
4                         & 25  & 600            &\textbf{171}  & 169&169&167 & 169      \\
4                         & 40  & 600            &148  & \textbf{155}&151&\textbf{155} & 155      \\
4                         & 60  & 600            &152  & 152 &152 &148 & 152      \\
4                         & 80  & 600            &140  & \textbf{141}&139 &\textbf{141} & 139      \\
4                         & 100 & 600            &136  & \textbf{138}&135 &136  &136     \\
4                         & 150 & 600            &128  & -  & \textbf{129}&\textbf{129}& 129      \\
4                         & 200 & 600            &\textbf{124} & - &122&123 & 123      \\ \hline
20                        & 10  & 600            &70  & 70&70&70  & 70       \\
20                        & 15  & 600            &62  & 62 &62&62 & 62       \\
20                        & 20  & 600            &54  & 54 &54&54  & 54       \\
20                        & 25  & 600            &51  & 51 &51&\textbf{52} & 52       \\
20                        & 40  & 600            &48  & 48 &\textbf{49} &\textbf{49} & 49       \\
20                        & 60  & 600            &46  & 46& 46&46 & 46       \\
20                        & 80  & 600            &43  & 43&43&43  & 43       \\
20                        & 100 & 600            &39  & 39&39 &\textbf{40} & 39       \\
20                        & 150 & 600            &37  & 37&37&37 & 37       \\
20                        & 200 & 600            &34  & - &34 &\textbf{35} & 35   
\\ \hline
Average                   & \multicolumn{2}{l}{} & 101.9 & - &102 & 101.8 &\textbf{102.4}  \\
\hline
\end{tabular}
\end{table}

Table~\ref{tb:rat} shows the results on the ``Rat'' dataset. The $GMPSUM$ does not have all the results due to the overflow. Moreover, $HH$ and $k_{analytic}$ produce the best results among all methods. The Rat dataset is almost uncorrelated. It is the real info of Rat's DNA and proteins (targets). For this type of dataset, $k_{analytic}$ shows a proper and higher performance in comparison to other non-$HH$ methods. However, the winner is the $HH$ with the highest performance among all methods for this dataset. It is necessary to mention that $GCoV$ has a lower performance than other methods. But, in some cases, the $GCoV$ could reach the best solution; we believe this depends on some properties of strings.

\begin{table}[H]
\centering
\caption{Comparison of state-of-the-art methods vs. our proposed methods on Virus dataset}
\label{tb:virus}
\begin{tabular}{llllllll}
\hline
\textbar{}$\Sigma$\textbar{} & N   & l              & $BS$-$Ex$\cite{djukanovic19}  & $GMPSUM$\cite{Nikolic21} &$k_{analytic}$ &$GCoV$ & HH  \\
\hline
4                         & 10  & 600            & 223 & \textbf{225}&223&218 & 223      \\
4                         & 15  & 600            & 202  &  203&203&\textbf{204} & 203      \\
4                         & 20  & 600            & 189  &  \textbf{191}&190&189  & 190      \\
4                         & 25  & 600            & 193 &  \textbf{194}&193&190  & 193      \\
4                         & 40  & 600            & 169 &  \textbf{170}&\textbf{170}&168 &170      \\
4                         & 60  & 600            & 166 &  \textbf{167}&166&162 & 166      \\
4                         & 80  & 600            & \textbf{162}  &  161&161&156 & 161      \\
4                         & 100 & 600            & 157 &  \textbf{158}&156&154 & 156      \\
4                         & 150 & 600            & 155 &  - &\textbf{156}&152  & 156      \\
4                         & 200 & 600            & 153 &  -&\textbf{155}&149 & 155      \\ \hline
20                        & 10  & 600            & 74  & 74&74&\textbf{75}  & 74      \\
20                        & 15  & 600            & \textbf{64} &  62&63&62  & 63       \\
20                        & 20  & 600            & 59 & 59&59&59  & 59       \\
20                        & 25  & 600            & 54 & 55&55 &55 & 55       \\
20                        & 40  & 600            & 49  & \textbf{51}&50&49  & 50     \\
20                        & 60  & 600            & 48  & 48&48&47   & 48       \\
20                        & 80  & 600            & 46  & 46&46&45  & 46       \\
20                        & 100 & 600            & 43  & 44&44&44  & 44       \\
20                        & 150 & 600            & 45  & 45&45&45  & 45       \\
20                        & 200 & 600            & \textbf{44}  & - &43&43  & 43       \\ \hline
Average                   & \multicolumn{2}{l}{} & 114.75 & - &\textbf{115} &113.3 & \textbf{115} \\
\hline
\end{tabular}
\end{table}

Table~\ref{tb:virus} states the results of the methods on the ``Virus'' dataset. This dataset contains almost uncorrelated strings of real virus genomes. The algorithms of $k_{analytic}$ and $HH$ have the best performance, the $BS$-$EX$ performs well, and finally, $GCoV$ has the worst performance among all of them.  

\begin{table}[H]
\centering
\caption{Comparison of state-of-the-art methods vs. our proposed methods on Random dataset}
\label{tb:rnd}
\begin{tabular}{llllllll}
\hline
\textbar{}$\Sigma$\textbar{} & N   & l    & $BS$-$Ex$\cite{djukanovic19}  & $GMPSUM$\cite{Nikolic21} & $k_{analytic}$ &$GCoV$ & HH  \\ \hline
4                         & 10  & 600     & \textbf{220}  & 218&\textbf{220}&217  & 220      \\
4                         & 15  & 600     & \textbf{203}  & 202&202&199 & 202      \\
4                         & 20  & 600     & 191  & 191&\textbf{192}&187  & 192      \\
4                         & 25  & 600     & 186  & \textbf{188}&187&183  &187      \\
4                         & 40  & 600     & 174  & 173 &\textbf{175}&170  & 175\\
4                         & 60  & 600     & \textbf{167}  & 165 &\textbf{167} &161  & 167      \\
4                         & 80  & 600     & 161  & 161&\textbf{162}&157  & 162      \\
4                         & 100 & 600     & 159  & 159&159&154 & 159      \\
4                         & 150 & 600     & \textbf{153}  & - &\textbf{153}&149  & 153      \\
4                         & 200 & 600     & 150  & - &\textbf{151}&147 & 151      \\ \hline
20                        & 10  & 600     & 62  & 62&62&61  & 62       \\
20                        & 15  & 600     & 51  & 51&51&51  & 51       \\
20                        & 20  & 600     & 47  & 47&47&47  & 47       \\
20                        & 25  & 600     & 44  & 44&44&44  & 44       \\
20                        & 40  & 600     & 38  & 38&\textbf{39}&38  & 39       \\
20                        & 60  & 600     & 35  & 35&35&35  & 35       \\
20                        & 80  & 600     & 33  & 33&33&33  & 33       \\
20                        & 100 & 600     & 32  & 32&32&32   & 32       \\
20                        & 150 & 600     & 29  & 29&29&29 & 29       \\
20                        & 200 & 600     & 27  & - &\textbf{28}&\textbf{28}  & \textbf{28}       \\ \hline
Average                   & \multicolumn{2}{l}{} & 108.1  & -&108.4&106.05 & \textbf{108.4} \\
\hline
\end{tabular}
\end{table}

Table~\ref{tb:rnd} declares that the $k_{analytic}$ outperforms the other methods for ``Random'' dataset. This dataset is completely uncorrelated.  

Also, we compare our methods with $BS$-$Ex$ and $GMPSUM$ on highly correlated datasets. The $k_{analytic}$ for the correlated datasets are based on Eq.~\ref{eq:nonuniformeq}. As mentioned before, the $BS$-$Ex$ method is not suitable for correlated strings, and it has the lowest performance in almost all cases of correlated strings. The first dataset is the $BB$ dataset which was introduced by Blum et al.~\cite{blum07}. The second one is a dataset that contains  $SARS$-$CoV$-$2$ genomes.
\begin{table}[H]
\centering
\caption{Comparison of state-of-the-art methods vs. our proposed methods on BB dataset}
\label{tb:bb}
\begin{tabular}{llllllll}
\hline
\textbar{}$\Sigma$\textbar{} & N   & l     &    $BS$-$Ex$\cite{djukanovic19}  & $GMPSUM$\cite{Nikolic21}&$k_{analytic}$ & $GCoV$ & HH  \\ \hline
2                         & 10  & 1000     &    604.8    & 629.7  & \textbf{635.7} &614.1&635.7   \\
2                         & 100  & 1000    &   531.3     & 556 & \textbf{559.4} &543.1&559.1    \\
4                         & 10  & 1000  &    424.4      &\textbf{467.5}  & 464.3&452.4&464.6     \\
4                         & 100  & 1000  &     324.7      & 364.5  & \textbf{365.6}&365.3&365.4  \\
8                         & 10  & 1000   &     289.9     & 343.4  & \textbf{359.6} &359.5&359.5     \\
8                         & 100  & 1000   &     201.5     & 244.1   & 242.1&\textbf{244.3}&243.3   \\
24                         & 10  & 1000   &      220.6    & 263  & 274.1&\textbf{277.1}&275.6      \\
24                         & 100 & 1000   &    102.2      & 128.9 & \textbf{130.8} &123&130.8     \\ \hline
Average                   & \multicolumn{2}{l}{} & 337.42 & 374.63 & 378.95&372.35&\textbf{379.25} \\
\hline
\end{tabular}
\end{table}
 Table~\ref{tb:bb} shows the result of the $BB$ dataset, in which we outperform $GMPSUM$ in 7 cases out of 8 cases, and the IP of the average solution for $k_{analytic}$ and $HH$ are, respectively, $1.15\%$, and $1.22\%$ for the $GMPSUM$ method. 

\begin{table}[H]
\centering
\caption{Comparison of state-of-the-art methods vs. our proposed methods on SARS-CoV-2 dataset}
\label{tb:sars_cov_2}
\begin{tabular}{llllllll}
\hline
\textbar{}$\Sigma$\textbar{} & N   & l     &    $BS$-$Ex$\cite{djukanovic19}  & $GMPSUM$\cite{Nikolic21} &$k_{analytic}$ & $GCoV$ & HH  \\ \hline
4                         & 10  & 400     &    182    & 189  & \textbf{198} &172&198   \\
4                         & 20  & 400    &   189     & \textbf{193} & 191 &176&191    \\
4                         & 30  & 400  &    173      &\textbf{178}  & 175 &156&175    \\
4                         & 40  & 400  &     150      & 164  & \textbf{168}&146&168  \\
4                         & 50  & 400   &     145     & 157  & \textbf{160}&141&160      \\
4                         & 60  & 400   &     139     & \textbf{153}   & \textbf{153}&139&153   \\
4                         & 70  & 400   &      137    & \textbf{148}  & 142 &136&142     \\
4                         & 80  & 400   &      140    & \textbf{151}  & 150 &140&150     \\
4                         & 90  & 400   &      150    & 158  & \textbf{163}&145&163      \\
4                         & 100  & 400   &      139    & 144  & \textbf{148}&132&148      \\
4                         & 110 & 400   &    \textbf{143}      & 141 & \textbf{143} &133&143     \\ \hline
Average                   & \multicolumn{2}{l}{} & 153.4 & 161.45 & \textbf{162.81} &146.9&\textbf{162.81}\\
\hline
\end{tabular}
\end{table}

Table~\ref{tb:sars_cov_2} shows the effectiveness of our methods on $SARS$-$CoV$-$2$ dataset. We obtain better results in 7 out of 11 cases, and the IP of the average solution is $0.84\%$.

The Fig.~\ref{fig:time} shows the running time of all algorithms on the four datasets. The results are presented on  those sets of strings with the number of alphabet $|\Sigma|=20$ with varying $N$. As the results on different datasets show, $k_{analytic}$ has the best run-time, and $GMPSUM$ has the worst run-time. The $HH$ method is a combination of $k_{analytic}$ and $GCoV$. As the plots demonstrate, when $HH$ are based on the former, its run-time is between $k_{analytic}$ and $GCoV$. But, when it is based on the $GCoV$, its run-time is higher than both of them. 

\section{Conclusion} \label{conc}
In this paper, we have worked on the Longest Common Subsequence problem. Most of the high-performance heuristics and algorithms to solve this problem are types of a tabular probabilistic scoring system utilized in a beam search schema. We have proposed a closed-form equation of the tabular recursive probabilistic computation in this work. We proved it and used it in our computations to have an improved analysis of the problem. We can estimate parameters effectively using the equation. Before that, all parameters are based on some guesses or suggestions without having an analytical reason.
Moreover, we propose two methods based on analytical interpretation. Our results show that our proposed methods outperform the state-of-the-art algorithms in both qualities of solutions and run-time. In future work, the closed-form equation can be used for better approximation with a lower computation overhead and time consumption. Moreover, it is possible to use the closed-form equations in palindromic strings for better analysis and consequently introduce improved heuristics.

\section*{Acknowledgement}
We thank Bahare Adeli for helping in designing some of the figures. We also thank Masih Hajsaeedi for proofreading the paper.
\bibliography{mybibfile}
\appendix

\section{Proof of \emph{Theorem 1}}\label{appendix1}
Let's assume the initial values of 
\[ 
p(k,n) = \alpha \times p(k - 1, n - 1) + \beta \times p(k, n - 1)
\]
are 
\[ 
p(k,n) = 0, k > n
\]
\[
p(k,n) = 1, k = 0
\]
We assume the correctness of the two following probabilities:
\[ 
p(k - 1,n-1) = 1 - \beta^{n-k+1} \left[ \sum_{i=0}^{k-2} \alpha^{i} {n-k+i \choose i} \right]
\]
\[ 
p(k,n-1) = 1 - \beta^{n-k} \left[ \sum_{i=0}^{k-1} \alpha^{i} {n-k-1+i \choose i} \right]
\]
Using induction, we have
\[ 
p(k,n) = \alpha p(k-1,n-1) + \beta p(k,n-1)
\]
\[ 
= \alpha \left( 1- \beta^{n-k+1} \left[ \sum_{i=0}^{k-2} \alpha^{i} {n-k+i \choose i} \right] \right) + \beta \left( 1- \beta^{n-k} \left[ \sum_{i=0}^{k-1} \alpha^{i} {n-k+i-1 \choose i-1} \right] \right)
\]
\[ 
= 1- \alpha \beta^{n-k+1}\left[ \sum_{i=0}^{k-2} \alpha^{i} {n-k+i \choose i} \right] - \beta^{n-k} \left[ \sum_{i=0}^{k-1} \alpha^{i} {n-k+i-1 \choose i-1} \right]
\]
\[ 
= \alpha + \beta -\left( \alpha \beta^{n-k+1}\left[ \sum_{i=0}^{k-2} \alpha^{i} {n-k+i \choose i} \right] + \beta \beta^{n-k} \left[ \sum_{i=0}^{k-1} \alpha^{i} {n-k+i-1 \choose i-1} \right] \right)
\]
\[
= 1 - \left( \beta^{n-k+1}\left[ \sum_{i=0}^{k-2} \alpha^{i+1} {n-k+i \choose i} \right] +  \beta^{n-k+1} \left[ \sum_{i=0}^{k-1} \alpha^{i} {n-k+i-1 \choose i-1} \right] \right)
\]
\[
= 1 - \left( \beta^{n-k+1}\left[ \sum_{i=1}^{k-1} \alpha^{i} {n-k+i-1 \choose i-1} \right] +  \beta^{n-k+1} \left[ \sum_{i=1}^{k-1} \alpha^{i} {n-k+i-1 \choose i} \right] + \beta^{n-k+1} \right)
\]
\[ 
= 1 - \left( \beta^{n-k+1} \left[ \sum_{i=1}^{k-1} \alpha^{i} {n-k+i-1 \choose i-1} + {n-k+i-1 \choose i} \right] + \beta^{n-k+1} \right)
\]
\[ 
= 1 - \left( \beta^{n-k+1} \left[ \sum_{i=1}^{k-1} \alpha^{i} {n-k+i \choose i} \right] + \beta^{n-k+1} \right)
\]
\[ 
= 1 - \left( \beta^{n-k+1} \left[ \sum_{i=0}^{k-1} \alpha^{i} {n-k+i \choose i} \right] \right) = p(k,n), QED.
\]
\section{Proof of \emph{Theorem 2}} \label{appndix2}
\[
p(k,n) = 1 - \beta^{n-k+1} \left[ \sum_{i=0}^{k-1} \alpha^{i} {n-k+i \choose i} \right]
\]
\[ 
= 1 - \beta^{n-k+1} \left[1 + \sum_{i=1}^{k-1} \alpha^{i} {n-k+i \choose i} \right]
\]
\[ 
= 1- \beta^{n-k+1} - \beta^{n-k+1} \sum_{i=1}^{k-1} \alpha^{i} {n-k+i \choose i}
\]
\[
= 1 - \beta^{n-k+1} - \beta \alpha \left[ \sum_{i=1}^{k-1} \frac{1}{i} Beta(\beta, n-k, i) \right], QED.
\]
\section{Comparison of $k_{analytic}$ with $k_{guess}$}
\label{appndix3}

\begin{table}[H]
\centering
\caption{Comparison of $k_{guess}$ and $k_{analytic}$ on Virus dataset}
\label{tb:t2}
\begin{tabular}{lllll}
\hline
\textbar{}$\Sigma$\textbar{} & N   & l             & $k_{guess}$\cite{mousavi12improved} & $k_{analytic}$
\\
\hline
4                         & 10  & 600         &223 & 223
\\
4                         & 15  & 600            &202  & \textbf{203}      \\
4                         & 20  & 600             & 188  & \textbf{190}      \\
4                         & 25  & 600            & 193 & 193      \\
4                         & 40  & 600          & 168 & \textbf{170}      \\
4                         & 60  & 600             & 165 & \textbf{166}      \\
4                         & 80  & 600             & 158& \textbf{161}      \\
4                         & 100 & 600             & \textbf{158}  &156     \\
4                         & 150 & 600            & 156  & 156      \\
4                         & 200 & 600             & 154  & \textbf{155}      \\ \hline
20                        & 10  & 600            & 74  & 74       \\
20                        & 15  & 600            & 62  & \textbf{63}       \\
20                        & 20  & 600            & 59  & 59       \\
20                        & 25  & 600            & 54  & \textbf{55}       \\
20                        & 40  & 600            & 49 & \textbf{50}       \\
20                        & 60  & 600            & 47 & \textbf{48}       \\
20                        & 80  & 600            & 45  & \textbf{46}       \\
20                        & 100 & 600            & 44 & 44       \\
20                        & 150 & 600            & 45 & 45       \\
20                        & 200 & 600            & \textbf{44}  & 43   
\\ \hline
Average                   & \multicolumn{2}{l}{} & 114.4 &\textbf{115}  \\
\hline
\end{tabular}
\end{table}
\begin{table}[H]
\centering
\caption{Comparison of $k_{guess}$ and $k_{analytic}$ on Random dataset}
\label{tb:t3}
\begin{tabular}{lllll}
\hline
\textbar{}$\Sigma$\textbar{} & N   & l           & $k_{guess}$\cite{mousavi12improved} & $k_{analytic}$ 
\\
\hline
4                         & 10  & 600          &214  & \textbf{220}      \\
4                         & 15  & 600            &\textbf{203}  & 202      \\
4                         & 20  & 600             & 191  & \textbf{192}      \\
4                         & 25  & 600            & 185 & \textbf{186}      \\
4                         & 40  & 600          & 172 & \textbf{175}      \\
4                         & 60  & 600             & 165 & \textbf{167}      \\
4                         & 80  & 600             & 161& \textbf{162}      \\
4                         & 100 & 600             & 158  &\textbf{159}     \\
4                         & 150 & 600            & 151  & \textbf{152}      \\
4                         & 200 & 600             & 150  & \textbf{151}      \\ \hline
20                        & 10  & 600             & 61  & \textbf{62}       \\
20                        & 15  & 600            & 51  & 51       \\
20                        & 20  & 600            & 47  & 47       \\
20                        & 25  & 600            & 44  & 44       \\
20                        & 40  & 600            & 38 & \textbf{39}       \\
20                        & 60  & 600            & 34 & \textbf{35}       \\
20                        & 80  & 600            & 32  & \textbf{33}       \\
20                        & 100 & 600            & 31 & \textbf{32}       \\
20                        & 150 & 600            & 29 & 29       \\
20                        & 200 & 600            & 28  & 28   
\\ \hline
Average                   & \multicolumn{2}{l}{} & 107.25 &\textbf{108.4}  \\
\hline
\end{tabular}
\end{table}

\begin{table}
\centering
\caption{Comparison of $k_{guess}$ and $k_{analytic}$ on BB dataset}
\label{tb:t4}
\begin{tabular}{lllll}
\hline
\textbar{}$\Sigma$\textbar{} & N   & l     &    $k_{guess}$\cite{mousavi12improved} & $k_{analytic}$  \\ \hline
2                         & 10  & 1000     &   626.3   & \textbf{635.7}   \\
2                         & 100  & 1000    &  554.2  & \textbf{559.4}     \\
4                         & 10  & 1000  &   456  & \textbf{464.3}     \\
4                         & 100  & 1000  &  359.7  & \textbf{365.6}  \\
8                         & 10  & 1000   &  352.8   & \textbf{359.6}      \\
8                         & 100  & 1000   &  236.9  & \textbf{242.1}   \\
24                         & 10  & 1000   & 267.3    & \textbf{274.1}      \\
24                         & 100 & 1000   &  128 &\textbf{130.8}      \\ \hline
Average                   & \multicolumn{2}{l}{} & 372.65 & \textbf{378.95} \\
\hline
\end{tabular}
\end{table}

\end{document}